\documentclass[apj]{emulateapj}

\newcommand{\new}[1]{#1}

\newcommand{\U}[1]{\ensuremath{\mathrm{~#1}}}
\newcommand{\kms}{\U{km\ s^{-1}}}
\newcommand{\yr}{\U{yr}}
\newcommand{\Myr}{\U{Myr}}

\newcommand{\pc}{\U{pc}}
\newcommand{\kpc}{\U{kpc}}
\newcommand{\Mpc}{\U{Mpc}}
\newcommand{\msun}{\U{M}_{\odot}}

\begin{document}
\title{$N$-body simulation of the Stephan's Quintet}
\accepted{13 September 2010}
\submitted{}

\author{Florent~Renaud\altaffilmark{1,2,3}, Philip~N.~Appleton\altaffilmark{1} and C.~Kevin~Xu\altaffilmark{1}}
\altaffiltext{1}{NASA Herschel Science Center, IPAC, California Institute of Technology,  MC 100-22, 770 S. Wilson Av., Pasadena CA 91125, USA}
\altaffiltext{2}{Observatoire de Strasbourg, 11 rue de l'Univerit\'e, 67000 Strasbourg, France}
\altaffiltext{3}{Institut f\"ur Astronomie der Univ. Wien, T\"urkenschanzstr. 17, A-1180 Vienna, Austria}
\email{florent.renaud@astro.unistra.fr}

%%%%%% UPDATE
%%% Diaz-Guimenez 2008
%%%%%%

%%%%%%%%%%%%%%%%%%%%%%%%%%%%%%%%%%%%%%%%%%%%%%%%%%\
\begin{abstract}

The evolution of compact groups of galaxies may represent one of the few places in the nearby universe in which massive galaxies are being forged through a complex set of processes involving tidal interaction, ram-pressure stripping, and perhaps finally ``dry-mergers'' of galaxies stripped of their cool gas.  Using collisionless $N$-body simulations, we propose a possible scenario for the formation of one of the best studied compact groups: Stephan's Quintet. We define a serial approach which allows us to consider the history of the group as sequence of galaxy-galaxy interactions seen as relatively separate events in time, but chained together in such a way as to provide a plausible scenario that ends in the current configuration of the galaxies. By covering a large set of parameters, we claim that it is very unlikely that both major tidal tails of the group have been created by the interaction between the main galaxy and a single intruder. We propose instead a scenario based on two satellites orbiting the main disk, plus the recent involvement of an additional interloper, coming from the background at high speed. This purely $N$-body study of the quintet will provide a parameter-space exploration of the basic dynamics of the group that can be used as a basis for a more sophisticated $N$-body/hydrodynamic study of the group that is necessary to explain the giant shock structure and other purely gaseous phenomena observed in both the cold, warm and hot gas in the group. 

\end{abstract}

%%%%%%%%%%%%%%%%%%%%%%%%%%%%%%%%%%%%%%%%%%%%
\section{Introduction}

\begin{figure*}
\plotone{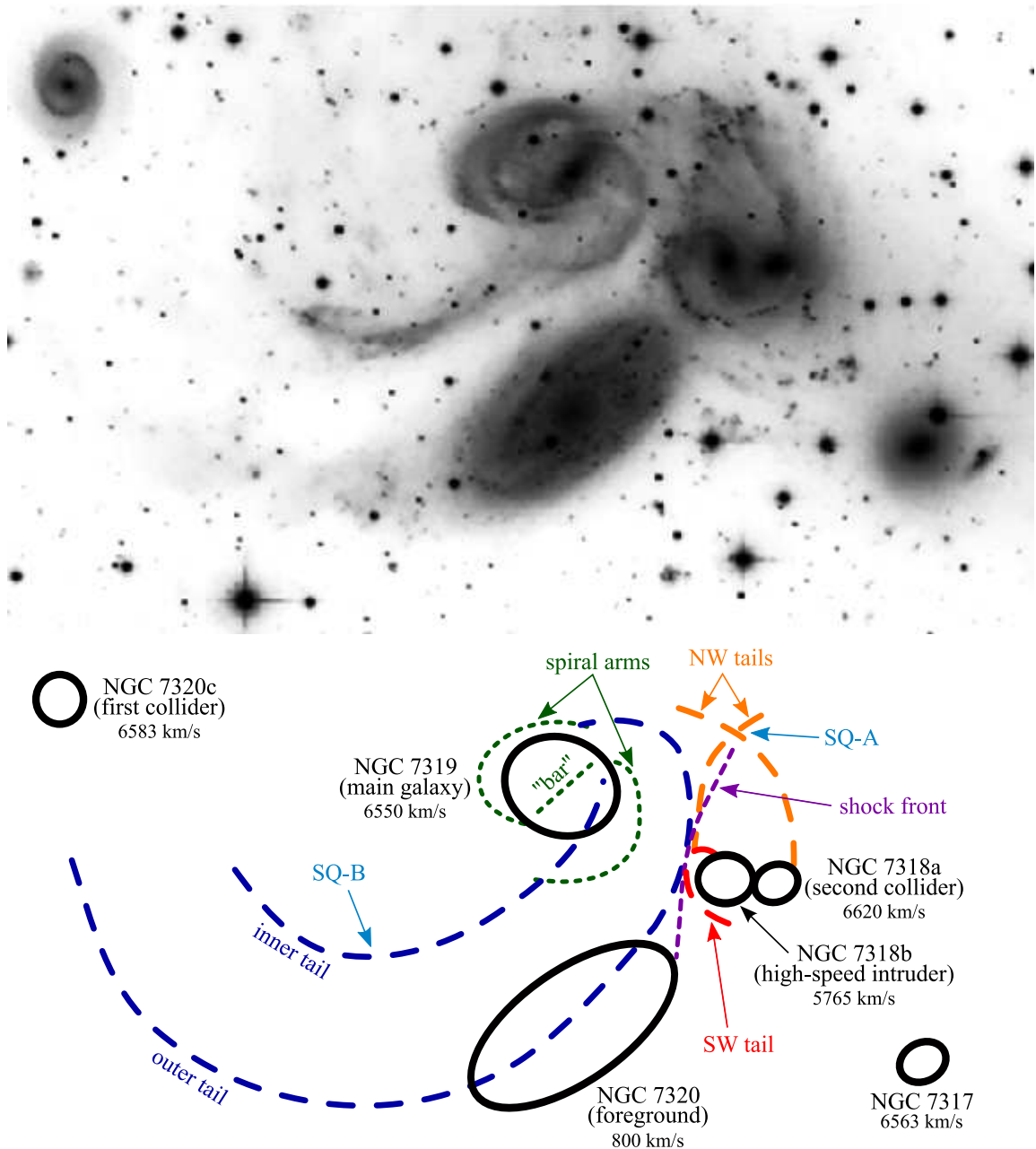}
\caption{Top: Stephan's Quintet members and the foreground galaxy, observed in the B-band. North is up, East is left. Image courtesy of Vicent Mart\'{i}nez, Fernando Ballesteros, Vicent Peris and Rodney Smith. Bottom: schematic view of the main morphological features of the group. Velocities of the galaxies are taken from \citet{Sulentic2001}.}
\label{fig:observations}
\end{figure*}

More than 130 years after its discovery \citep{Stephan1877}, the Stephan's Quintet (aka HCG~92 or Arp~319; hereafter SQ, see Figure~\ref{fig:observations}) remains enigmatic, despite being one of the most well studied compact groups in the local universe. One of the particularities of SQ is the wide range of redshifts of its supposed members. The southern galaxy (NGC~7320) has a velocity of $800 \kms$, i.e. about $5000 \kms$ less than the rest of the group \citep{Burbidge1961}. Although early hypotheses argued that this galaxy was a case of existence of non-Doppler redshift \citep{Arp1973a}, it is more likely a foreground galaxy, as there are no evidences of interaction with other members of the quintet. This has been confirmed by redshift-independent distances measurements  \citep{Shostak1974a, Shostak1974b, Allen1980, Moles1997}, and by the clear resolution of individual stars in NGC~7320 by the recently refurbished Hubble Space Telescope \citep[ERO4]{Noll2008}. However, the higher-redshift system can rightfully be classified as a quintet because of the discovery of an additional group member, NGC~7320c, which clearly shows optical links with NGC~7319 and likely was responsible for at least one major tidal filament in the group as suggested by \citet{Arp1973b}. Finally, one of the members, NGC~7318b appears to have recently fallen into the group from the background and is currently interacting with the intragroup medium (IGM).

The puzzling aspects of Stephan's Quintet do not end with its kinematics. It has long been known that its cool gas distribution is very peculiar. Almost all of the \ion{H}{1} emission is seen outside of the optical body of the group members \citep{Allen1980, Shostak1984, Williams1999, Williams2002}, and even its cold molecular gas is mainly displaced from the galaxies and lies in apparently isolated clumps in the IGM, with only a few concentration within the optical structure of the largest galaxy NGC~7319. 

Even more peculiar was the discovery of a radio continuum filament lying between the NGC~7318b/a pair and NGC~7319 \citep{Allen1972, vanderHulst1981, Xu2003}, perhaps suggesting a shock compression caused by the interaction of the infalling galaxy NGC~7318b with the group. The filament was subsequently found to be a strong emitter of X-rays \citep{Pietsch1997, Trinchieri2003, Trinchieri2005, OSullivan2009}, and optical spectroscopy provided more evidences that the structure was a giant intergalactic shock wave \citep{Xu2001, Xu2003}. Surprisingly, the X-ray luminosity in the shock structure is exceeded by emission from lines of warm molecular hydrogen seen in the Mid-IR \citep{Appleton2006, Cluver2010}. All of these peculiarities seem to relate to a complex and rich history of collisions and interactions in the group which has led to a highly disturbed state for both the gas and stars in this system \citep[e.g.][]{Moles1997, Sulentic2001, Guillard2009}. Attempting to unravel the complex past history of the group via simulations is the main aim of the present paper.

\new{Our understanding of the dynamical properties of compact groups in general \citep[e.g.][]{Barnes1985, Diaferio1994, Athanassoula1997, Aceves2002} relies on statistical explorations covering broad ranges of parameters \citep{McConnachie2008, DiazGimenez2008}. Although this numerical and theoretical approach provides an important global description for the interaction process, it misses a comparison with compact groups actually observed. Therefore, while the physics involved in the evolution of such objects is well-known, the question of scenarios that might bring a new light on observations of individual groups remains poorly addressed.}

In this contribution, we investigate the possible chains of events that may lead to the observed configuration of SQ. Because of the intrinsic challenge that represents the simulation of the group, we concentrate first on the collisionless dynamics, leaving the additional level of complexity of the hydrodynamics to a forthcoming paper.

We first draw a quick picture of the group, listing the features we use as clues to understand the formation history. In Section 3, we propose an original approach to explain the web of features of SQ. We distinguish two main regions in the quintet and describe the formation scenarios used in Section 4. We close with a discussion and a summary of our main results.

%%%%%%%%%%%%%%%%%
\section{Morphological features}
Early observations showed that \ion{H}{1} is widely spread outside the optical regions of SQ \citep{Allen1980, Shostak1984}. One classical explanation is that the gas has been stripped-off the main galaxies of the group during tidal interactions and close passages. More recent measures \citep{Williams2002} confirmed that the radial velocities of \ion{H}{1} regions are in good agreement with this idea, as they match the velocity of their former host galaxies.

Optical images show two large tidal arms, both extending from NGC~7319 to the southern region of the group (in blue on Figure~\ref{fig:observations}). The outer one, very diffuse, passes behind the foreground galaxy (NGC~7320). Another arm, more concentrated, is almost parallel to the first one and seems to point to the small NGC~7320c galaxy. The differences in length and luminosity of these two structures have been explained by two separated interaction steps: the outer arm is produced by a first passage of an intruder close to NGC~7319 while the inner plume formed in a more recent event \citep{Moles1997}. This would explain that the second tail is still very young and not quite extended (regarding to the other one). \new{\citet{Xu2005} pointed out that it took $> 5\times 10^8 \yr$, after the close encounter with NGC~7319, for NGC~7320c to move to its current position. They argued that it is likely that the new tail is triggered by a close encounter between NGC~7319 and NGC~7318a, which is $\sim 3$ times closer to NGC~7319 than NGC~7320c is.}

The other important region is the pair of galaxies (NGC~7318a and NGC~7318b) on the west side of the group. The smallest one (NGC~7318b) has a velocity of $5765 \kms$ whereas the velocity of its counterpart (NGC~7318a) is closer to the rest of the group \citep[$6620 \kms$,][]{Sulentic2001}. \citet{Xu2003, Xu2005} interpreted this difference as a shift of the first galaxy $\sim 100 \kpc$ toward us, after a head-on collision of the pair members $\sim 10^7-10^8 \yr$ ago. This scenario explains the peculiar \ion{H}{1} distribution outside the optical disk. \new{Although the observational data do not permit to conclude on the origin of the filaments at the North of the pair (called NW tails, orange on Figure~\ref{fig:observations}), we note that these structures seem to expand in a cylindrical symmetric way, as opposed to the usual tidal tails that are confined to a well-defined direction with respect to their host galaxy. This ``ring-like'' structure observed in SQ backups the hypothesis of a head-on collision between the two members of the pair \citep{Higdon1995}.} As mentioned before, the recent collision of the intruder with the IGM triggered a large scale shock, between the pair and the main galaxy (purple on Figure~\ref{fig:observations}).

Note also that the combination of internal and tidal processes shaped smaller pseudo-spiral arms in the main galaxy and the intruder (green and red on Figure~\ref{fig:observations}). The question of their origin is still to be address but it is likely that their formation has been triggered by tidal interactions.

These tidal and gaseous structures observed today reveal a complex gravitational history. Therefore, in addition to observations, numerical simulations are a good way to test hypotheses and reproduce the evolution of such a complex system. The morphology and kinematics of the structures presented here and on Figure~\ref{fig:observations} are the main clues in our understanding of the quintet, and thus, the starting point of our modeling study.

%%%%%%%%%%%%%%%%%%%%%%%%%%%%%%%%%%%%%%%%%%%%
\section{Serial approach}
The most natural way to grasp such a complex object is to set global initial conditions. In this case, the simulation starts using an arbitrary time, since one can consider the progenitors as isolated galaxies. Thus, the game is to find the orbital parameters that lead to an interaction. This method has been widely used for the simulations of pairs or simple groups \citep[e.g.][]{Toomre1972, Barnes1988, Hibbard1995, Dubinski1996, Barnes2004, Renaud2008}. Indeed, it is rather easy to create individual models for each galaxy, to set the phase-space separations between them, and let them evolve, following the implemented laws of physics. However, dynamical friction and rapid evolution of the morphology add some difficulties to this exercise and a lot of attempts are often required before finding a good match with observational data.

Another approach consists in exploring the parameters space with restricted simulations and select the best match among them \citep{Barnes2009}. The very efficient simulation of point-mass galaxies surrounded by test particles \citep[see e.g.][]{Toomre1972} allows to cover a broad range of parameters. \citet{Theis2001} combined this method with genetic algorithms. Thanks to a maximum likelihood estimation, a good match with an observational data set can be rapidly found. However, such a method faces problems to properly describe the orbital decay, an important characteristic of galaxy mergers, in particular for compact groups and thus, is not applicable here.

In the classical $N$-body approach, before interaction, it is still possible to approximate the trajectories of the centers of mass with a two-body solution. However, close passages would induce a loss of energy (dynamical friction) and mass (creation of the tails), that will invalidate the two-body result. Therefore, when a third galaxy enters the already affected system, its orbit (i.e. impact point and relative velocity) strongly differs from those one could expect analytically. That is why, for this study which implies complex trajectories, we use the traditional try-and-see method to reproduce the quintet.

As SQ is currently at a typical intermediate stage of its evolution (and thus, displays a very complex web of features), it is very difficult to define a complete set of the numerical parameters which best describes the observed object. Therefore, we decided to follow a serial (as opposed to parallel) approach, assuming that most of the features of the quintet have been created separately and then, gathered without strong interaction or disturbance between them. This quite bold assumption allows us to split the problem into simpler interacting pairs, and to get a good idea of the historical evolution of compact groups in general. Our method first creates a galaxy-galaxy interaction and uses the result as one of the progenitor of the second step, and so-on.

Luckily, SQ does lend itself to such a slicing in both spatial and time dimensions. Previous studies \citep{Moles1997,Xu2005} suggested that the eastern tidal tails were built $\sim 500-700 \Myr$ ago for the longest one and $\sim 200 \Myr$ ago for the narrowest one, although the western pair of galaxies would have an interaction history of $\sim 10 \Myr$ only.

Therefore, our serial approach consists in creating the two main tidal arms in the eastern region, involving NGC~7319 (the main galaxy), NGC~7320c (first collider) and likely NGC~7318a (second collider), and then add NGC~7318b (the intruder) in the western side for the last interaction event. Because of its separation with the rest of the group and the lack of linking tails, we assume that the first interaction of the remaining galaxy (NGC~7317) has not taken place yet. It may fall into the center of the group in the future but is not directly linked to it at present time.

\begin{figure}
\plotone{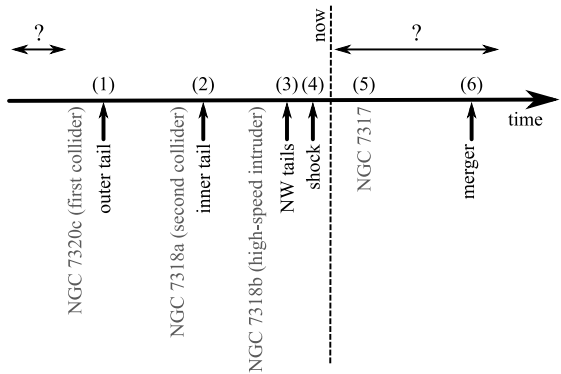}
\caption{Schematic view of the major steps in the history of SQ. Very early phases and future (marked with $?$) are subject to arbitrary assumptions that may not be realistic. See Section~\ref{sec:discussion} however, for a discussion on the possible future of the group.}
\label{fig:time_scheme}
\end{figure}

The interaction scenario, divided into six major steps, is schematically presented in Figure~\ref{fig:time_scheme}: (1) formation of the outer tidal tail, (2) creation of the inner (younger) tail, (3) western interaction between NGC~7318b and the rest of the group, (4) large scale shock, (5) involvement of the fifth galaxy (NGC~7317) and finally, (6) possible merger. According to this scheme, the currently observed objects stand at the end of step (4). We do not have many clues about the last two steps but we can reasonably assume that SQ will fit in the general evolution theory of compact groups, as described in \citet{Johnson2007}, i.e. tidal harassment of its gaseous envelop and, \new{possibly}, formation of a massive ellipsoid from the progenitors \new{(see a discussion on the shape of merger remnants in \citealt{Weil1996}). Depending on the relative velocities of the galaxies, the final merger would involve all or only part of the members, and could occur in more than a Hubble time, as suggested by \citealt{Mamon1987} (see also \citealt{Athanassoula1997}, and \citealt{Aceves2002} for a discussion on the overmerging issue of compact groups).} It is likely that a similar approach applies to other compact groups, which undergo a sequence of collisions possibly leading to the hierarchical formation of a massive galaxy.

From a simulation standpoint, the last step of gathering them into a single model is much more difficult to perform, even with a good description of the interactions of pairs of galaxies. Because the progenitors cannot be considered in an equilibrium stage after they have experienced their first interaction, one has to carefully re-tune the initial parameters to get similar results than in isolation. NGC~7320c is already in a remote position when NGC~7318b enters the group, and thus should not be affected by the intruder. NGC~7318a however, stands very closely to the main galaxy. The trajectory of NGC~7318b must be re-set to compensate the attraction of NGC~7319 and keep the same impact point on the disk of NGC~7318a. In addition to this geometrical effect, one also has to consider the mass loss of the second collider from which a tidal arm has already been stripped. Differences with the simulation of the simple pair should be limited however, thanks to the high velocity of the intruder: NGC~7318b undergoes the perturbations of distant galaxies for a very short time, which reduces their effects.

%%%%%%%%%%%%%%%%%%%%%%%%%%%%%%%%%%%%%%%%%%%%
\section{Modelling Stephan's Quintet}

%%%%%%%%%%%%%%%
\subsection{Numerical method}
For this study, we use a purely gravitational $N$-body code to reproduce a global scheme of the interacting history of SQ. Hydrodynamical approach will be added in a forthcoming work\footnote{We assume that the gas, which represents only a small fraction of the total mass, does not strongly affect the morphology of the system.}. Each spiral galaxy is set up as self-gravitating disk + bulge + dark matter halo model using the {\tt magalie} software \citep{Boily2001}, included within the Nemo stellar dynamics package \citep{Teuben1995}. We use extended bulge components, considering that they account for both the compact central bulge and the spheroid. The tree code {\tt gyrfalcON} \citep{Dehnen2002} solves the equations of motion, with a Plummer softening kernel ($\epsilon = 23 \pc$) \new{and a tolerance parameter $\theta=0.6$}. \new{The conservation of energy has been verified by running a simulation of the model of NGC~7319 in isolation. We noted a relative difference of $10^{-4}$ after $600 \Myr$. This small value gives us confidence in the results obtained with this code and this set of parameters.}

All our progenitors, except NGC~7320c, are made of an exponential disk, a \citet{Hernquist1990} bulge and a truncated isothermal sphere as dark matter halo. The first collider (NGC~7320c) does not show a clear internal tidal signature, hence we assume a spherically symmetric distribution and \new{only use the bulge and halo components}. The mass ratio between all individual particles is set to unity, to avoid any two-body heating. The scaling of the galaxies in mass, velocity and length is in a such way that the average density and the virial parameter are conserved. Table \ref{tab:param} lists the parameters used for our best model. Considering that the initial mass (dark matter included) of the main galaxy (NGC~7319) is about $\sim 2.5 \times 10^{12} \msun$, our mass resolution is $\sim 3 \times 10^6 \msun$. In the following, $t = 0$ corresponds to the snapshot that best-matches the observations.

\begin{deluxetable*}{lcccc}
\tablecaption{Parameters of the simulation.\label{tab:param}}
\tablehead{\colhead{Parameter} & \colhead{NGC~7319} & \colhead{NGC~7320c} & \colhead{NGC~7318a} & \colhead{NGC~7318b}\\ & (main galaxy) & (1st collider) & (2nd collider) & (high-speed intruder)}
\startdata
Number of particles	& 870,000 & 22,000 & 300,000 & 150,000\\
\hline
\multicolumn{5}{c}{Disk parameters} \\
Number of particles	& 108,700 & 0 & 37,500 & 18,700\\
Mass			& 1.00 & 0.00 & 0.34 & 0.18\\
Vertical scalelength	& 0.10 & - & 0.07 & 0.06\\
Radial scalelength	& 1.0 & - & 0.7 & 0.6\\
Cut-off radius		& 10.0 & - & 7.0 & 5.6\\
Toomre parameter	& 1.5 & - & 1.5 & 1.5\\
\hline
\multicolumn{5}{c}{Bulge parameters}\\
Number of particles	& 54,400 & 1,600 & 18,700 & 9,400\\
Mass			& 0.50 & 0.10 & 0.17 & 0.09\\
Scalelength		& 2.0 & 0.7 & 1.4 & 1.1\\
Cut-off radius		& 10.0 & - & 7.0 & 5.6\\
\hline
\multicolumn{5}{c}{Dark matter halo parameters} \\
Number of particles	& 706,900 & 20,400 & 243,800 & 121,900\\
Mass			& 6.50 & 1.30 & 2.21 & 1.17\\
Scalelength		& 2.0 & 0.7 & 1.4 & 1.1\\
Cut-off radius		& 20.0 & - & 14.0 & 11.2\\
\hline
\multicolumn{5}{c}{Disks inclinations} \\
($\theta_x$,$\theta_y$) & (0,0) & - & (180,23) & (0,-23)\\
\hline
\multicolumn{5}{c}{Initial coordinates} \\
$(x,y,z)$ & (0.0,0.0,0.0) & (12.2,6.4,12.2) & (-50.9,0.0,23.7) & (-8.6,10.8,-133.0)\\
$(v_x,v_y,v_z)$ & (0.0,0.0,0.0) & (-0.80,0.30,-0.50) & (0.60,-0.06,-0.12) & (0.04,0.00,1.00)
\enddata
\tablecomments{The parameters are given in numerical units ($G=1$). The conversion factors are $3.2\times 10^{11} \msun$, $2.3 \kpc$, $775 \kms$ and $2.9 \Myr$. The non-dimensionalization is based on the separation between NGC~7319 and NGC~7318a measured by \citet{Lisenfeld2004} and the velocities given by \citet{Sulentic2001}, assuming a recession velocity of $6400 \kms$ and $H_0 = 75 \U{km\ s^{-1}\ Mpc^{-1}}$ which gives a distance of $85 \Mpc$.}
\end{deluxetable*}

%%%%%%%%%%%%%%%%%
\subsection{Formation of the large scale tidal tails}
Since \citet{Toomre1972}, it is well-known that the tidal features formed during an interaction are made of the material stripped from the galactic disks, because they are dynamically cooler than the bulges and the halos. \ion{H}{1} maps by \citet{Williams2002} suggest that the two major arms in SQ originate from the main galaxy (NGC~7319), as the greatest velocity difference between its disk and the tips of both arms is $\sim 0.9 \%$ only. The elongation of the two arms clearly shows that the causes of their formation must be prograde (direct) interactions between collider(s) and a very large disk.

%%%%%%%
\subsubsection{Outer tail}
The outer tail, which spans from the main galaxy to the south and then curves toward the east (see Figure~\ref{fig:observations}), is obviously the oldest tidal structure in the quintet. Indeed, its very low surface brightness and large width suggest that it had time to expand and vanish into the IGM. Furthermore, the tidal stripping of the associated \ion{H}{1} distribution (noted Arc-S by \citet{Williams2002}, see their Figure 5) by the collider implies that no major interaction had expelled the gas before, and thus that this event is the first in the history of the group. As the \ion{H}{1} tail points toward \new{the North, close to} NGC~7320c, it is very likely that this galaxy is responsible for the first interaction (and thus justifies its nickname: the first collider). Both the \ion{H}{1} distribution and the low brightness of the tail suggest that it has formed in a distant interaction, during which the gas has been stripped off while only the less bound stars were tidally affected.

\begin{figure}
\plotone{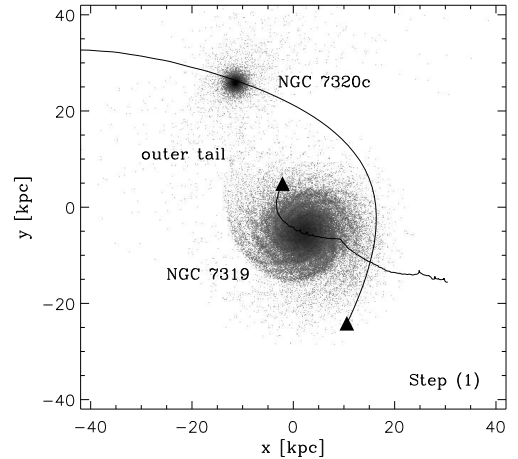}
\caption{Morphology of NGC~7319 and NGC~7320c after the first interaction ($t=-260 \Myr$). The outer tail is very faint for our stellar model but should be much more visible with a gaseous component (similarly to the observations of \ion{H}{1}). All the parameters used (including the orientation of the line of sight) are those of the final model of the entire quintet. \new{The solid lines are the orbits of the center of mass of the two galaxies, with triangles marking their positions before the interaction.}}
\label{fig:1920c}
\end{figure}

In our modeling approach, this first step accounts for the creation of a long tidal tail originating from the main disk, a small counter-tail, and an orbit for the collider that drives it toward its current position. We note that, because of its mainly gaseous structure, the tail must count only few stellar particles of our model. To do so, we set NGC~7320c on a highly eccentric prograde orbit around NGC~7319, with a non-penetrating encounter, as shown in Figure~\ref{fig:1920c}. This setup has to bring it in the vicinity of the main disk but keep enough energy to send it away, at $\sim 85 \kpc$ from NGC~7319, as observed.

There is no clear clue about the inclination of the orbital plane with respect to the plane of the sky. The two progenitor galaxies show similar line of sight (LOS) velocities, which suggests \new{a weak (or even no) proper motion along this line, by contrast with NGC~7318b. Therefore, it is likely that both orbital planes are nearly identical.} Note that the repartition of the measured LOS velocity between recession and proper motion of a given galaxy is unknown. If one assume a common group velocity (based on the one of the main galaxy), it seems reasonable to attribute the difference to both a geometrical offset along the LOS and a proper motion. In the case of NGC~7319 and NGC~7320c, the velocity difference is too small ($\sim 30 \kms$) to allow any conclusion. Therefore, we first set the orbit of the first collider on the plane of the sky.

In the Keplerian approximation (valid until the pericenter passage), the main galaxy stands at one of the two foci of the orbit of the collider. Because NGC~7319 must undergo a prograde encounter, the direction of rotation of NGC~7320c along the orbit is determined by the spin of NGC~7319. Here again the choice is arbitrary because of the lack of observational clues. We set the rotation of the main galaxy to be counterclockwise, when viewed from the Earth, implying the same direction for the trajectory of the collider. Therefore, to bring it to its observed position, it must pass the main disk westward.

%%%%%%%
\subsubsection{Inner tail}
The material of the two tails clearly shares the same origin: the main disk. However, the inner tail shows a much higher surface brightness, is thiner and shorter, which suggests that it has been created after the outer one, and thus during a second interaction. A glance at the velocities and \ion{H}{1} distribution of the members of SQ limits the possibilities to NGC~7320c and NGC~7318a.

\begin{figure}
\plotone{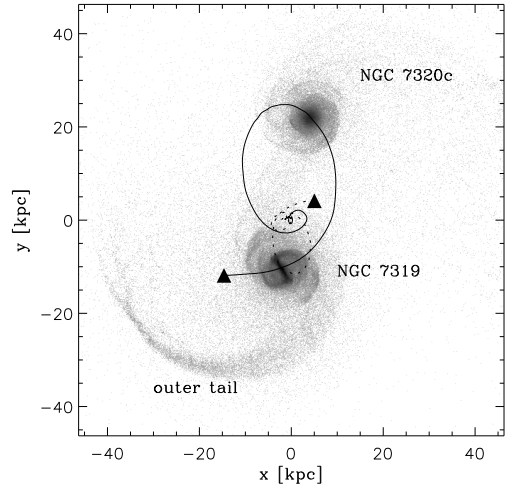}
\caption{Hypothesis of the creation of two tails by a single collider. The figure plots the morphology of the system after the first pericenter passage. The outer tail is already well developed in the southern part. Solid and dotted lines shows the orbital decay in the trajectories of the collider and the main galaxy, respectively. Initial positions are marked with triangles. The second pericenter passage will lead to a strong loss of orbital energy and finally the merger of the two progenitors. Therefore, this scenario has been rejected.}
\label{fig:twopassages}
\end{figure}

In the first case, NGC~7320c has to pass twice in the vicinity of the main galaxy: once to create the outer tail, and a second time to build in inner one. This constrains the possible trajectories to bound orbits, with enough energy to ``escape'' from the gravitational potential of main disk twice and finally go to the observed remote position ($\sim 85 \kpc$ away from NGC~7319). However, in order to strip such a amount of material (\ion{H}{1} and stellar tidal tails), the collider must pass very close to the center of the main disk, where the density is high. In these regions, the dynamical friction plays an important role and forbids the satellite galaxy to keep the same orbital energy. Because the friction already makes it brake after the first pericenter passage, the orbital decay after two passages does not allow the required separation, as visible on Figure~\ref{fig:twopassages}. To verify this, we have run a systematic survey on the orbital parameters space, made of 96 simulations involving NGC~7319 and NGC~7320c. As no model gives the right final separation between the two progenitors after two passages, we conclude that it is very unlikely that the same galaxy creates both arms during two different interaction events.

\begin{figure}
\plotone{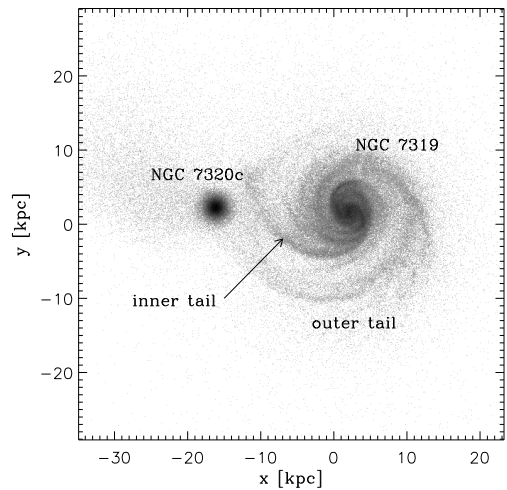}
\caption{Hypothesis of the creation of the two tails with one passage only. Both structures have correct positions and inclination. However, the age difference between the plumes is not reproduced by this scenario.}
\label{fig:m51}
\end{figure}

In another scenario, the collider and the main disk interact only once, and both tails are created at the same time. Indeed, the effect of the dynamical friction is considerably reduced and the orbital decay becomes less important than in the two passages scenario. With a highly inclined encounter \citep[an approach commonly used for M~51, see e.g.][]{Toomre1973, Theis2003} the tidal bridge forms between the collider and the main disk, while a counter-tail is created in the opposite side of the disk, representing the inner and outer plumes, respectively. In our simulations, the mass ratio, the inclination of the orbital plane and the impact velocity have been adjusted to avoid the destruction of the disk on the one hand, and the merger of the two progenitors on the other. After few tests, a good agreement with observations has been found for the position and inclination of the tails, as shown in Figure~\ref{fig:m51}. However this scenario fails to take into account the age difference between the two plumes (because both are created at the same time), and the \emph{a priori} low inclination of the orbital plane with respect to the plane of the sky. In addition, both plumes present similar surface brightness when formed the same way, another difference with the observations, suggesting that this scenario is not correct.

The last remaining possibility involves NGC~7318a. When considering this galaxy as the second collider, the age difference ($\sim 3-5 \times 10^8 \yr$) between the two tails can easily be explained as a result of two well separated events. As the ``two events - single collider'' scenario reveals itself to transfer too much of its orbital angular momentum through dynamical friction, we set two collider to hit the main disk sequentially. Because the outer tail is more extended and older, the cause of its creation had more time to escape the central region of the group. Therefore, NGC~7320c is likely responsible for it, while the inner tail would be created by NGC~7318a, few $10^8 \yr$ after. In this case, the outer tail is a bridge between the first collider and the main disk, while the inner tail is a classical arm (or counter-tail) from the interaction with the second collider.

The numerical simulation of this scenario is more difficult than usual, as the second interaction has to occur in an already disturbed system, without affecting the existing structures (i.e. mainly the outer tail). Any Keplerian approximation (even before the pericenter passage) is not possible in this pseudo three-body problem. In addition, the existence of extended tidal structures voids the point-mass approach. For simplicity however, we set the initial conditions with Keplerian orbits, keeping in mind that the actual trajectory will rapidly differ from those of our naive setup. In the previous section, we have set the spin of NGC~7319 to be counterclockwise. To get the prograde encounter for the main galaxy, the second collider must hit NGC~7319 southward before going toward the west. Unfortunately, the observational data reveal an extremely dense and complex web of gaseous and stellar structures between these two galaxies (recall Figure~\ref{fig:observations}). Hence, it is rather difficult to look for tidal signatures for NCG~7318a, that may indicate if its collision with the main disk is prograde or retrograde. However, a diffuse structure linking the two galaxies is visible in deep images \new{(see e.g. the HST/WFC3 image by \citealt{Noll2008}, ERO4)} and thus, suggests the presence of a tidal bridge from NGC~7318a toward NGC~7319. Note that most of these tidal features have been strongly disturbed by the recent collision with NGC~7318b (see the next Section).

\begin{figure}
\plotone{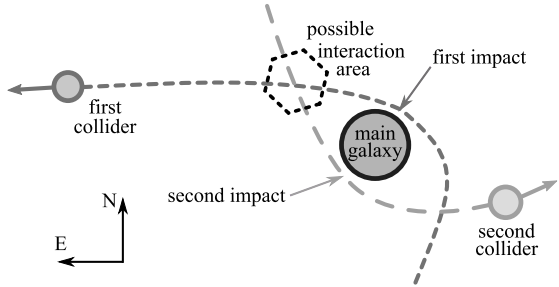}
\caption{Schematic view of the orbits of the two colliders around the main galaxy. According to the ages of the tails (and thus the ages of the tidal interactions) the two colliders might interact where their trajectories cross. Circles show the position of the three galaxies at present time.}
\label{fig:tails_scheme}
\end{figure}

According to their current positions, the two colliders have trajectories which could make them interact, between the two collision events with the main galaxy. Indeed, after its collision, NGC~7320c flies eastward, north of the main disk, while NGC~7318a is approaching to circle around the the main galaxy from north-east to south-west, as shown in Figure~\ref{fig:tails_scheme}. The lack of observational clues (like tails between NGC~7318a and NGC~7320c or a common halo) suggests that the two galaxies have not interacted with each other. In order to keep the same timescales and the same orbits with respect to the main galaxy, the only solution to avoid the collision between NGC~7318a and NGC~7320c is that they do not share the same orbital plane. According to the measured radial velocities, the second collider might be behind the first one, at present time. This suggests to tilt the orbital plane of NGC~7319a so that it starts closer to the Earth, interacts with the main galaxy, and then goes behind. By doing this, the trajectories of the two colliders do not cross.

\begin{figure}
\plotone{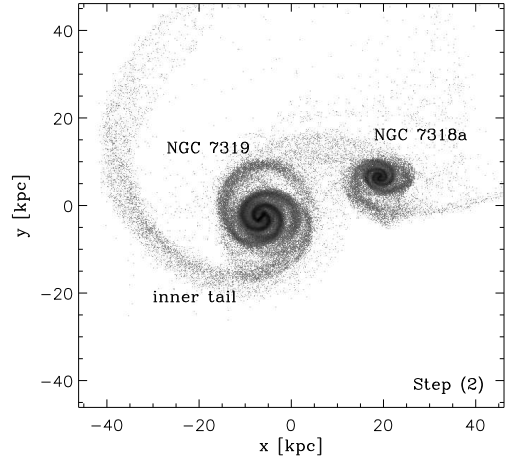}
\caption{Morphology of NGC~7319 and NGC~7318a after the formation of the inner tail ($t=-80 \Myr$). NGC~7320c is included in the simulation, but not shown here. The tail and the spread ISM between the two galaxies are clearly visible. A secondary tail originating from NGC~7318a is expanding toward the west but will vanish during the next steps of the interaction history of SQ. \new{The outer tail, created by NGC~7320c,  is already too difused to be visible.}}
\label{fig:1918a}
\end{figure}

This scenario involving three galaxies allows to tune the initial parameters in order to place the two intruders at their right positions after the interaction, i.e. in the far east for NGC~7320c and near the center, ready for a collision with NGC~7318b, in the case of NGC~7318a. The inclination of the orbital axis and the initial starting point along the trajectories have been chosen so that the two progenitors reach their observed position at the same time (i.e. now). The initial Keplerian orbit of the second collider has to be corrected to account for the mass loss and shift in position due to early collision. The final result is shown in Figure~\ref{fig:1918a}

We note the presence of a strong spiral and even a barred shape in NGC~7319 (see Figure~\ref{fig:observations}). Previous studies suggested that such structures could result from a tidal interaction \citep{Walker1996, Berentzen2004}. On the one hand, it is very likely that more recent interactions erased or at least disturbed them. On the other hand, deep images clearly reveal that such asymmetric structures still exist in the quintet, in particular in the main disk. Our simulation managed to reproduce the spiral arms and even a weak bar after the second collision. They are transient features that will vanish within a few dynamical times. \new{By integrating NGC~7319 in isolation during $600 \Myr$, we have checked that these structures are not intrinsic to the model, but created by the interaction with the other galaxies.}

%%%%%%%%%%%%%%%%%
\subsection{Formation of the western region}

For the next step in the interaction history of the quintet, NGC~7318b hits the triplet NGC~7318a-19-20c and its already existing web of tidal features. Although the velocity field reveals itself to be extremely complex in this region \citep{Sulentic2001, Williams2002}, one can extract some hints from it. First of all, the large velocity difference measured between NGC~7318a and NGC~7318b ($\sim 850 \kms$) clearly requires an orbit highly inclined from the plane of the sky for the NGC~7318b (the intruder). Secondly, because the internal tidal structures (spirals and bars) are still blurry, the interaction has taken place recently. Therefore, the velocity of NGC~7318b should be high enough to separate it from the rest of the group in such a small amount of time. Consequently, it is possible to consider the intruder as a galaxy coming from the background at high velocity and hitting the group in a configuration close to the one currently observed. 

%%%%%%%
\subsubsection{Tidal tails of the pair}
\begin{figure}
\plotone{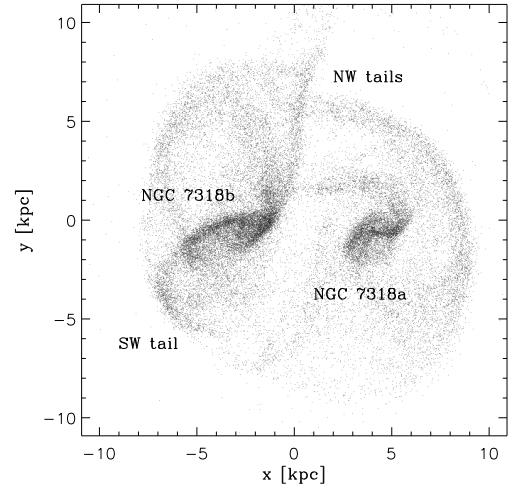}
\caption{Simulation of the western pair, in isolation. Here, the crossing of the two northern tails is real (i.e. not a projection effect). However, the main tail of NGC~7318b is not correctly positioned with respect to its nucleus. The crossing of the tails with a curvature toward the counterpart galaxy \new{(and not away from it, as shown here)}, as observed in SQ, is very unlikely in a classical encounter. Hence these structures should have a different origin.}
\label{fig:pair}
\end{figure}

The most visible features in this region are two thin tails (NW tails in Figure~\ref{fig:observations}) expanding northward from the two galaxies NGC~7318a/b. Their mean measured velocity of $\sim 6000 \kms$ suggests that they stand between the two galaxies. Therefore, these structures have been created by the tidal interaction of the second collider with the high speed intruder. It seems that each tail originates from one of two galaxies, the western one being from NGC~7318a while the other starts from NGC~7318b. Simulations of encounters show that a tidal structure either links the two galaxies (bridge) or point away from the counterpart (classical tail). Figure~\ref{fig:pair} shows the result of an interaction between two disks galaxies that leads to similar morphology than the pair. However, the northern tail of NGC~7318b is connected to its disk on the right-hand side, contrary to the observations of SQ. Hence the crossing of the two NW tails we see in SQ would be a projection effect along the line of sight. In other words, the two tails would not share the same plane. However, this crossing matches the position of the starburst region SQ-A \citep{Xu1999}. It is likely that the over-density due to the physical overlap of the tails favors the formation of stars. Therefore, the crossing is not a simple projection effect but a volume common to both tails which have formed, in this case, in an unusual way. 

NGC~7318b shows another thin arm (SW in Figure~\ref{fig:observations}) containing tidal dwarf galaxies candidates \citep{Hunsberger1996, Mendes2001}. The origin of this structure pointing toward NGC~7318a remains mysterious, which confirms the complex history of this region of the group.

A possible explanation consists in considering all these arms as a single \new{ring-like} structure. Face-on collisions of disk galaxies create such rings \citep{Higdon1995} and also explain the star formation along it \citep{Beirao2009}. The velocity field measurements in this area \citep{Williams2002} place the ring between the two galaxies, but closer to NGC~7318b, suggesting that the external features come from its progenitor disk and underwent a significant brake regarding to the more compact nucleus. However, this scenario does not explain the asymmetry of this structure nor the crossing of two NW tails in the SQ-A area. For the moment, the observations reveal extremely complex morphology and velocity field whose dynamical explanation seems still out-of-reach. However, it is certain that this region results from a recent high-velocity collision between a disk intruder coming from the background and NGC~7318a, which had already been altered by a previous interaction with the main galaxy. Consequently, one could expect the progenitor NGC~7318b to have a disk large enough to form the ring-like structure (or to contain all the material that has been spread away).

\begin{figure}
\plotone{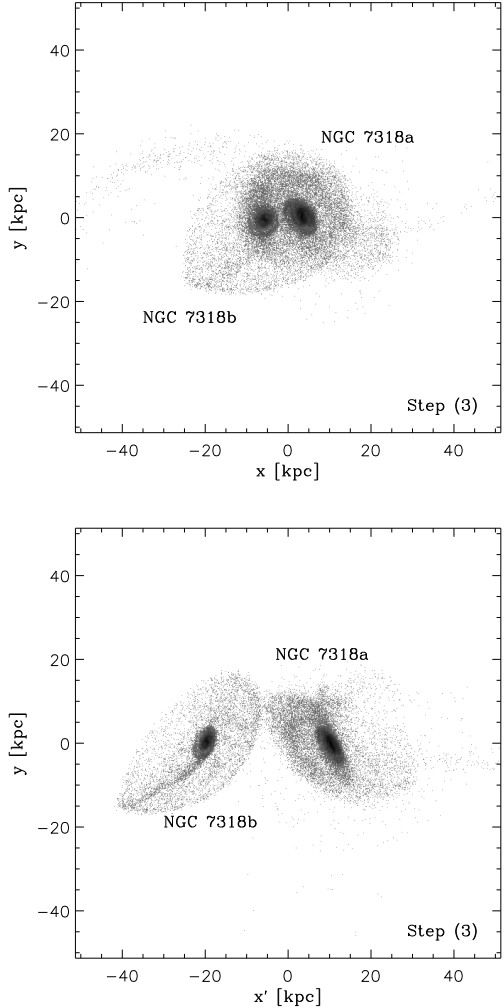}
\caption{Morphology of the western pair shown in the plane of the sky (top panel) and after a rotation of $30^\circ$ around the vertical axis (bottom panel). Again, the other galaxies (the first collider and the main disk) are included in the simulation but not shown here. \new{The tidal structure on NGC~7318b's disk exhibit a circular configuration, as opposed to a classical tail. This} is well-visible on the bottom panel, with a spoke running southward. Such spoke could explain the SW-tail observed.}
\label{fig:18a18b}
\end{figure}

Our goals in the modeling of this region remains modest, as we simply want to create two northern arms and possibly a third tail, southern. The fine details of the tidal structures, the crossing and the internal bar or spirals in NGC~7318b are out of the scope of the present paper. The results are visible in Figure~\ref{fig:18a18b}.

Tests showed that the required velocity corresponds to an unbound orbit \new{of NGC~7318b with respect to the rest of the group}, even when taking dynamical friction into account. Indeed, considering that the inclination of the disk of NGC~7318a is small with respect to the line of sight, the intruder undergoes a strong friction during only the short time it needs to go through this disk. In addition, the density of the IGM within which the intruder flies is low and hence, the dynamical friction there does not play a crucial role. In this view, the high-velocity intruder does not loose enough of its orbital energy to remain in the gravitational well of the group, and escapes in the foreground. If this scenario is confirmed, it might imply a different origin for the intruder than for the other galaxies in the group. Note however that these questions remain highly speculative and will not be addressed in this paper.

%%%%%%%
\subsubsection{Large scale shock}
The other major feature of the western part of SQ is the $\sim 40\kpc$ long shock, between the high-speed intruder and the main galaxy. Previous studies suggested that the shock results from the high-speed collision between NGC~7318b and the IGM \citep{Pietsch1997, Trinchieri2005, Xu2003, Appleton2006}. In the scenario presented here, this medium comes from the interactions involving NGC~7319, NGC~7320c and NGC~7318a. It contains the gaseous and stellar components tidally stripped during the formation of the eastern tails and the ``bridge'' between the main galaxy and its second collider. Indeed, observations in several wavelengths reveal that this region is filled with atomic and molecular gas \citep[see e.g.][and references therein]{Lisenfeld2002, OSullivan2009}. The recent intrusion of NGC~7318b in this medium triggered a hydrodynamic shock that propagates eastward. In addition to the relatively high densities of the tidal tails, the shock might also help to trigger the star formation, as observed in SQ-A, via radiative transfer \citep[see the suggestion of][and his application to the Mice]{Barnes2004}.

Note that, because of its tidal origin, this gas stands between NGC~7319 and NGC~7318a (which is behind the main galaxy). Considering that the high-speed intruder comes from the background, it has hit NGC~7318a before the gaseous medium (which is closer to us). In this respect, the shock is younger than the tidal features observed in the western region (NW and SW tails). It is clear that the origin, formation and propagation of the shock result from hydrodynamic processes, not discussed here. Hence, the shock is not modeled in the present work, but will be in forthcoming studies.

%%%%%%%%%%%%%%%%%
\subsection{Final model}

\begin{figure*}
\plotone{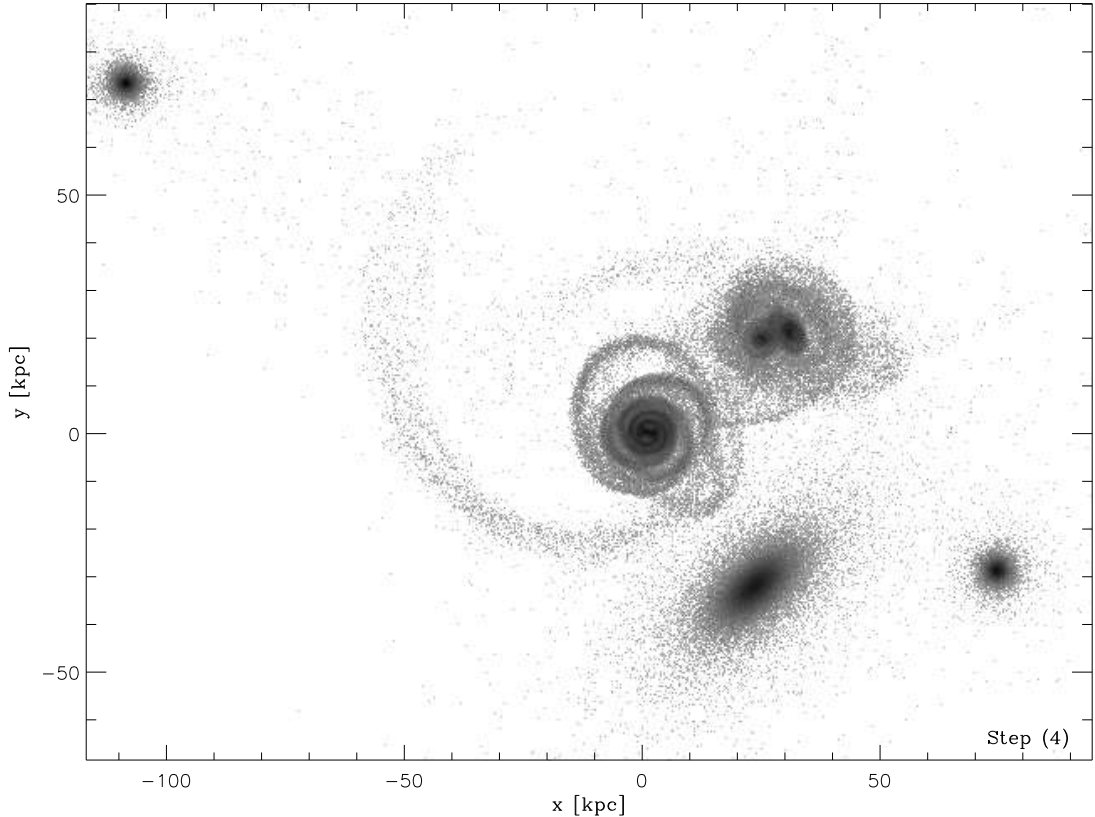}
\caption{Morphology of the simulated Stephan's Quintet, at $t=0$. This figure stacks the previous Figures~\ref{fig:1920c}, \ref{fig:1918a} and \ref{fig:18a18b}, this time showing all the galaxies. The additional galaxies NGC~7317 (west) and NGC~7320 (foreground) have been manually overlaid to the configuration of the rest of group, for graphical purposes only. Note that these two galaxies are \emph{not} included in the simulation itself, because of lack of observational signatures of past interactions that may constrain their trajectories and velocities.}
\label{fig:model}
\end{figure*}

In Figure~\ref{fig:model}, we show the complete simulation of the group, at $t=0$. Even if small-scale details are not perfectly reproduced, the large-scale configuration of the galaxies and the tidal features match well the observations. We note that the bar and the spiral pattern in NGC~7319 are induced by external effects (the interaction with NGC~7320c and NGC~7318a) and not internal processes. Most of the IGM is stripped off the main disk but also originates from the late collision between NGC~7318a and NGC~7318b.

\begin{figure}
\plotone{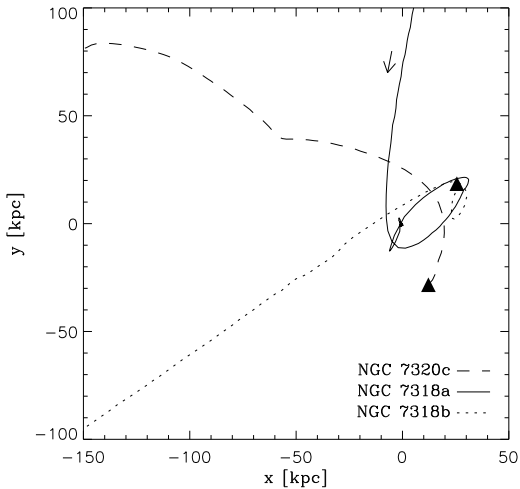}
\caption{Orbits of the first collider (dashed), second collider (solid) and the high-speed intruder (dotted) in the plane of the sky, with respect to the position of the main galaxy. \new{Initial positions are marked with triangles. The arrow indicates the direction of NGC~7318a, whose initial position is out of the frame of the Figure}. The change of curvature on the orbit of NGC~7320c reflect the modification of the central potential of the group, due to interactions. The merger of NGC~7318a is clearly visible after its second pericenter passage.}
\label{fig:orbits}
\end{figure}

Figure~\ref{fig:orbits} displays the orbits of the three satellites in the reference frame of the main galaxy. Before its interaction with the group, NGC~7318b has an almost radial motion and does not show a large displacement in the plane of sky. However, after the collision, the attraction of the two massive galaxies (NGC~7319 and NGC~7318a) deviates the intruder from its initial orbit. This explains the peculiar shape of its trajectory, in projection.

As in the observations, the outer plume is very faint (and almost not visible) in our simulations. Still, the external part of the main disk has been spread toward NGC~7320c, in a very diffuse fashion. \new{Almost all the stars of this tail visible in Figure~\ref{fig:model} are between NGC~7320c and the tip of the inner plume. A comparison with \ion{H}{1} observations suggests indeed that this tail would be made of mainly the gaseous envelop of the main galaxy and thus, would not count a lot of stars. New runs including hydrodynamics are currently performed to confirm this idea \citep{Hwang2010}. Note that for more of the stellar component to be ejected, the impact parameter should be smaller, but the main galaxy would be affected too strongly to permit the creation of the second, inner tail.}

Our model fails to place the western pair of galaxies at its observed position, i.e. southern than in our simulation. Note that this problem cannot be corrected by simply changing the orientation of the orbits, because this will imply a rotation of the tidal tails too. An inclination of the orbital plane, leading to projection effects could be a solution, although no observational element indicates such a configuration.

The fine structures in the west pair are not well reproduced. Indeed, a description of such features requires a very precise setting of the parameters. Unfortunately, NGC~7318a is not in equilibrium (after its encounter with the main galaxy) when it hits the intruder\new{, as many transient features have not dissolved yet}. Therefore the impact point can not be precisely defined, nor the relative velocity or the mass ratio. In addition, most of the structures observed seems to have an important (even dominant) hydrodynamic part, that cannot be reproduced in purely gravitational simulations. The asymmetry of the system and the over-densities matching the positions of the NW-tails are correctly reproduce however, thanks to the formation of the ring-like structure.

\begin{figure*}
\plotone{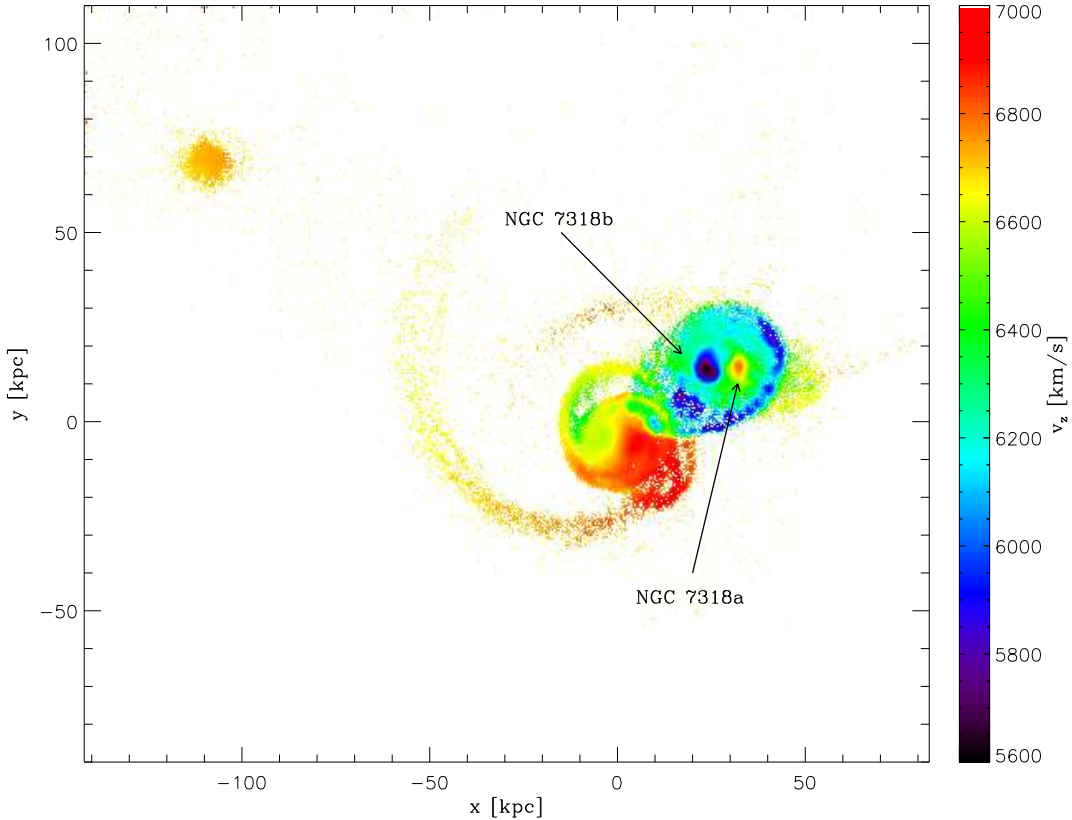}
\caption{Velocity along the line of sight, of the final model. A systemic velocity of $6650 \kms$ (yellow) has been assumed. The shift of the high-speed intruder with respect to the rest of the group, and in particular NGC~7318a is clearly visible. The \new{ring-like} strucutre also contrasts well with it enviroment.}
\label{fig:velocity}
\end{figure*}

In addition to the morphology, one has to check that the kinematics of the group is also well-reproduced. Figure~\ref{fig:velocity} displays the velocity along the line of sight for the galaxies included in the simulation. \new{If one assumes that the kinematics of the gas follows that of the stars, it becomes possible to compare the velocity patterns existing in our model to those measured from \ion{H}{1} data in previous studies.} All the features show velocities close to the ones measured, in particular the gradient between the high-speed intruder and the second collider. The velocity pattern of the \new{NW and SW tails} also matches the data of \citet[their Figures 10, 11 and 12]{Williams2002}. 
 
%%%%%%%%%%%%%%%%%%%%%%%%%%%%%%%%%%%%%%%%%%%%
\section{Discussion}
\label{sec:discussion}

%%%%%%%%%%%%%%%%%
\subsection{Assumptions of this model}

Despite a number of observational data about SQ, we have made some assumptions for the simulation presented above. First of all, the distribution of dark matter is not well constrained by the observations. Here, we consider one spherical halo per galaxy. Note however, that a more complex distribution involving triaxial cuspy galactic halos \citep[see e.g][]{Hayashi2007} and even a global group halo is possible. Changing the dark matter distribution in the simulation will require a fine tuning of the orbital parameters of the galaxies because the halos, being the most massive and gravitationally dominant components of the progenitors, set the details of their dynamical evolution. However, this would not affect the chain of events described above, nor the role of each galaxy.

Similarly, we have assumed that the rotation of NGC~7319 is counterclockwise. Because its disk lies in a plane close to those of the sky, the rotation curve is not measurable, which makes our hypothesis purely arbitrary. Indeed, if one considers a clockwise rotation instead, the orbits of the two colliders simply need to be flipped, in order to keep prograde encounters. In this case, NGC~7320c passes southward of the main galaxy, while NGC~7318a goes northward (corresponding to a $180^\circ$ flip of Figure~\ref{fig:tails_scheme} around its horizontal axis). By simply changing the initial conditions, it is possible to reach a comparable final configuration, which shows that the model presented here is not unique. The sequence of events and the origins of the structures are, however, very likely to follow our dynamical scenario.

Finally, the mass ratio between the progenitors is not strongly constrained. In our model, we set it thanks to the relative luminosities of the galaxies (assuming a constant mass-to-light ratio) and then tuned it according to the dynamical results obtained during our exploration of the parameters space. One can reasonably assume that NGC~7318a and NGC~7318b\new{, as observed today, have similar masses (ratio close to 1:1). Taking into account a  mass loss of $~30\%$ of the second collider during its interaction with the main disk, one obtains that NGC~7318a must be initially more massive than the high-speed intruder.}

%%%%%%%%%%%%%%%%%
\subsection{Future of SQ}

\begin{figure}
\plotone{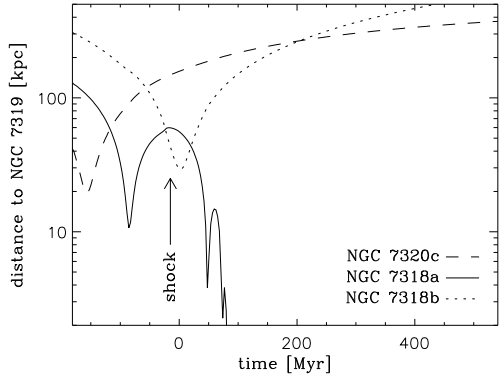}
\caption{Evolution of the 3D distance between SQ members and its main galaxy NGC~7319, in log-scale. Dashed, solid and dotted lines correspond to NGC~7320c, NGC~7318a and NGC~7318b respectively. This figure shows the sequential scenario of the creation of SQ, as all the episodes of the interaction history are clearly visible and well-separated in time. \new{The arrow marks the approximative time of the shock between NGC~7318b and the IGM}. Again, $t=0$ corresponds to the best-match to the observations.}
\label{fig:distance}
\end{figure}

The sequential splitting of events matches the spatial distribution of the galaxies, in the sense that interactions are well separated in time and space. Figure~\ref{fig:distance} plots the evolution of the distance between each galaxy and those of NGC~7319. This clearly shows that the interaction events (i.e. the low distances) follow a sequence in time. It also allows to estimate the age of structures like, e.g. the large-scale shock, once a triggering event is assumed, \new{by searching for, e.g. a minimum distance between the progenitors}. We note that our values \new{($\sim 150 \Myr$ for the age of the outer tail, $\sim 80 \Myr$ for the young one, and about $10 \Myr$ for the shock)} match well the estimations from the literature presented earlier.

Figure~\ref{fig:distance} shows that NGC~7318a merges with the main galaxy at $t \approx 100 \Myr$. Simple estimates of escape velocities reveal that \new{the mass of all the simulated galaxies but the intruder (3 × 1012 M⊙) corresponds to an escape velocity of $\sim 250 \kms$ at a distance of $200 \kpc$, i.e. less than the estimated velocity of the intruder. This is confirmed by our simulation: NGC~7318b is unbound with respect to the rest of the group. Note that Figure~\ref{fig:distance} suggests that this is the case of NGC~7320c too.} However, the remaining galaxy NGC~7317 (not simulated here) does not yield a high LOS velocity and is close to the central region of the group. Therefore, it will likely merge with the main galaxy and thus modify the gravitational potential of the group. This may change the fate of the escaping galaxies, by slowing them down and maybe making them fall back into the group core. \new{Following our estimates, a central mass of $\sim 2\times 10^{13} \msun$ is required to bind NGC~7318b which yields a LOS relative velocity of $850 \kms$ in both the observations and our model.} Unfortunately, the lack of tidal signatures and estimates for the tangential velocity of the missing galaxy forbids us to go further than speculations.

%%%%%%%%%%%%%%%%%%%%%%%%%%%%%%%%%%%%%%%%%%%%
\section{Conclusion and perspectives}

Using collisionless $N$-body simulations, we studied some possible formation scenarios of the Stephan's Quintet. We have run more than 3,000 attempts to simulate a fraction or the entire quintet and constructed a dynamical scenario based on the separation in time and space of the different interaction events. This scenario can be summarized as follow:
\begin{itemize}
\item mild interaction between the first collider (NGC~7320c) and the main disk (NGC~7319). The old, faint tidal tail is created. \new{Because it is difficult to see in our simulation, it may mainly consist} of the \ion{H}{1} external envelop of the main galaxy.
\item second interaction, between NGC~7318a and NGC~7319. The younger tail is built and points away from the second collider. A significant mass fraction \new{($\sim 30\%$ of the disks)} of both galaxies is expelled into the IGM. Internal structures are created in the main disk. 
\item frontal collision of the high-speed intruder (NGC~7318b), coming from the background, with the second collider (NGC~7318a). Several arm-like tidal features are created, possibly via the formation of a \new{ring-like structure}.
\item interaction between the high-speed intruder and the IGM. A large scale shock front forms.
\item interaction between NGC~7317 and the rest of he group. This event has not taken place yet and is thus not implemented in our simulations.
\end{itemize}

After an exhaustive exploration of the parameters space, it appeared that other hypotheses for the formation of the tidal tails are very unlikely. \new{In particular, the origin and relative age of the outer and inner tails seem to be very well constrained. The case of the western structures should be investigated further when including hydrodynamics aspects.}

Such a scenario clearly shows a splitting of interaction events in space and time, \new{over $\sim 100 \kpc$ and a few $\sim 10^8 \yr$,} a property which may characterize the evolution of compact groups. In the special case of SQ, the configuration observed reveals past episodes, whose imprints are the tidal tails, current events like the collision of the western pair of galaxies, and future interactions e.g. between NGC~7317 and the rest of the group. In slightly more evolved groups, each event would be characterized by a tidal signature whose age may denote this time sequence. However, at the end of their hierarchical evolution, the light of compact group is mostly dominated by the central galaxy, which makes diffuse plumes more difficult to detect.

Future numerical studies will explore the response of a gaseous component to the multiple interactions of SQ \citep{Hwang2010}. This might bring new clues about the formation and early evolution of the large-scale shock. With a detailed hydrodynamical model coupled to our gravitational description, we could be able to study the large-scale shock, the starburst regions, and furthermore, to answer a lot of questions raised by a growing number of observations over a broad range of wavelengths.

%%%%%%%%%%
\acknowledgments
We acknowledge the anonymous referee for very constructive comments that helped to improve this work. It is a pleasure to thank Curt Struck, Jeong-Sun Hwang, Pierre-Alain Duc and Ute Lisenfled for interesting discussions, and Christian Boily and Christian Theis for their support during this work.

\end{document}